# Evolution of ferromagnetic circular dichroism coincident with magnetization and anomalous Hall effect in Co-doped rutile $TiO_2$

H. Toyosaki, T. Fukumura,[1] Y. Yamada, and M. Kawasaki

*Institute for Materials Research, Tohoku University, Sendai 980-8577, Japan*

Magnetic circular dichroism (MCD) of rutile $Ti_{1-x}Co_xO_{2-\delta}$ is systematically examined with various $x$ and $\delta$ to reveal a phase diagram for the appearance of ferromagnetism at higher carrier concentration and Co content. The phase diagram exactly matches with that determined from anomalous Hall effect (AHE). The magnetic field dependence of MCD also shows good coincidence with those of the magnetization and AHE. The coincidence of these independent measurements strongly suggests single and intrinsic ferromagnetic origin.

PACS number(s): 75.50.Pp; 78.20.Ls; 72.25.Dc; 72.25.-b

[1] Electronic mail: fukumura@imr.tohoku.ac.jp

Ferromagnetic semiconductors provide an arena for spintronics devices utilizing both charge and spin degrees of freedom of charge carriers.[1] The ferromagnetism can be controlled by electrical or optical stimulus as demonstrated for III-V ferromagnetic semiconductors,[2,3] but the operation has been limited to temperatures far below room temperature. Therefore, extensive researches have been devoted to search for higher Curie temperature ($T_C$) ferromagnetic semiconductors. Oxide semiconductor family is one of the most promising host compounds to achieve higher $T_C$.[4,5] Indeed, room temperature ferromagnetism was revealed in anatase and rutile $Ti_{1-x}Co_xO_{2-\delta}$.[6,7] For rutile $Ti_{1-x}Co_xO_{2-\delta}$, anomalous Hall effect (AHE) was shown to emerge above certain thresholds of the electron concentration and Co content. The anomalous Hall conductivity $\sigma_{xy}^{AHE}$ scaled with the longitudinal conductivity $\sigma_{xx}$ as $\sigma_{xy}^{AHE} \propto \sigma_{xx}^{1.5\text{-}1.7}$.[8]

Magneto-optical effect of ferromagnetic semiconductor and AHE are ferromagnetic responses of the photo-excited and dc-driven charge carriers, respectively. Magneto-optical effect and AHE reflect the off-diagonal part of the dielectric tensor, corresponding to the nonzero and zero components of angular momentum, respectively. Among various magneto-optical effects, magnetic circular dichroism (MCD) spectrum is proportional to the energy derivative of the absorption coefficient ($\alpha$) spectrum,[9] hence, the absorption threshold should have a good coincidence with MCD spectrum. We have already reported on the ferromagnetic MCD varying systematically on $x$ and $\delta$ for anatase $Ti_{1-x}Co_xO_{2-\delta}$.[10] Rutile $Ti_{1-x}Co_xO_{2-\delta}$ was reported to show ferromagnetic MCD,[11] although the dependence on the carrier concentration ($n$) and $x$ was not investigated systematically. In this paper, we report on the systematic evolution of ferromagnetic MCD in rutile $Ti_{1-x}Co_xO_{2-\delta}$ with increasing $n$ and $x$, for identical samples used in Ref. 8 so as to make a direct comparison between MCD and AHE.

Rutile $Ti_{1-x}Co_xO_{2-\delta}$ epitaxial thin films were fabricated by laser molecular beam epitaxy.[8] $Ti_{1-x}Co_xO_2$ ceramics targets were ablated by focused KrF excimer laser pulses. The films with (101) orientation were grown on *r*-sapphire substrates at 400 °C. By varying the oxygen pressure during the growth ($P_{O2}$) from $1 \times 10^{-7}$ Torr to $1 \times 10^{-4}$ Torr, $\delta$ and consequently $n$ can be tuned systematically. The reflection high energy electron diffraction intensity was monitored *in situ* during the growth. The intensity oscillation was observed at initial stage of the growth, and the surface morphology of the resulting films (70-120 nm thick) showed atomically flat surfaces consisted of the steps and terraces observed by atomic force microscopy. Segregation of secondary phase was not observed under careful inspections by x-ray diffraction, atomic force microscopy, scanning electron microscopy, and transmission electron microscopy. The absorption spectra in visible to ultraviolet region were deduced from the transmittance and reflectance measurements. MCD was measured with the magnetic field normal to the film plane (Faraday configuration).[10] The magnetization was also measured in magnetic field normal to the film plane with a superconducting quantum interference device magnetometer. The Hall effect in magnetic field normal to the film plane was measured for photolithographically patterned Hall bars. $n$ was deduced from the ordinary part of the Hall resistance, that is the linear slope component with respect to the magnetic field. The mobility was approximately the same for all the conducting samples (~0.1 $cm^2$/Vs).

Figure 1 shows the absorption and MCD spectra measured at room temperature for the $Ti_{0.97}Co_{0.03}O_{2-\delta}$ with different $n$. The films are mostly transparent in visible region, and $\alpha$ increases from ~3 eV as shown in Fig. 1(a). Only for $n = 1 \times 10^{22}$ $cm^{-3}$ sample, the absorption threshold significantly shifts to higher energy side and the in-gap absorption develops around 1.5 eV. These features are similar to the cases for reduced bulk $TiO_{2-\delta}$.[12] As shown in Fig. 1(b), MCD signal is negligible for lower $n$ ($7 \times 10^{18}$ $cm^{-3}$, $4 \times 10^{19}$ $cm^{-3}$), whereas the large signals appears for higher $n$ ($2 \times 10^{20}$ $cm^{-3}$, $1 \times 10^{22}$ $cm^{-3}$). The latter spectra are rather broad: MCD is negatively large from 1.5 eV to about the absorption threshold, and turns to be positive for the higher energy region. The in-gap MCD might be caused by the strong spin-orbit coupling.[13] For $n = 1 \times 10^{22}$ $cm^{-3}$ sample, both the absorption and MCD spectra show the similar amount of blue shift. The inset in Fig. 1(b) shows the MCD spectra for $n = 2 \times 10^{20}$ $cm^{-3}$ sample measured in various magnetic fields. MCD at any photon energy changes in the same manner with magnetic field and saturates in higher magnetic field, indicating that the ferromagnetic MCD is originated from the single ferromagnetic source.[14]

Figure 2 shows the $x$ dependence of MCD spectra ((a) and (b)) and magnetic field response ((c) and (d)) for conducting $Ti_{1-x}Co_xO_{2-\delta}$ samples (grown in different $P_{O2} = 10^{-6}$, $10^{-7}$ Torr). The MCD signal is negligible for $x = 0$ and $x = 0.01$. From $x = 0.03$, the MCD signal appears and increases with increasing $x$. The magnetic field dependence of MCD signal apparently shows the ferromagnetic behavior for $x = 0.03$-$0.10$: the steep increase in low magnetic field and the clear saturation in high magnetic field. The hysteresis is hardly seen in this magnetic field scale. The saturation field depends on $P_{O2}$: $P_{O2} = 10^{-6}$ Torr samples show steeper increase of MCD than that for $P_{O2} = 10^{-7}$ Torr samples

The present study reveals that the appearance of ferromagnetic MCD depends both on $n$ and $x$ as mapped out in Fig. 3, where solid and open symbols represent ferromagnetic and paramagnetic MCD responses, respectively. The phase boundary as seen in Fig. 3 is exactly consistent with that for AHE previously reported.[8]

Figure 4 shows the magnetic field dependences of the magnetization, MCD and Hall resistivity for $Ti_{0.90}Co_{0.10}O_{2-\delta}$ with (a) $n = 2 \times 10^{20}$ cm$^{-3}$ and (b) $n = 4 \times 10^{21}$ cm$^{-3}$ measured at room temperature. The applied magnetic field is perpendicular to the film plane for all the measurements. The curves coincide each other, expect for the non-essential deviation of the Hall resistivity in higher magnetic field due to the ordinary Hall effect. The magnetization, MCD, and Hall resistivity represent the ferromagnetic response of the static spins, photo-excited carriers, and itinerant charge carriers, respectively. Therefore, the coincidence of these measurements strongly suggests the presence of the single ferromagnetic source. Accordingly, we conclude that the ferromagnetism is not caused by insulating or optically inactive ferromagnetic precipitation. We also note that $x$ is much below the percolation limit, hence, the magnetic Co cluster should not take significant role for the appearance of AHE.[15,16] If we plot $\sigma_{xx}$ and $\sigma_{xy}^{AHE}$ values available in Ref. 14 and 15, the all date falls in the same trend in Fig. 4 in Ref. 8.

In conclusion, rutile $Ti_{1-x}Co_xO_{2-\delta}$ is shown to have ferromagnetic MCD for higher $n$ and $x$. The good coincidence is demonstrated not only in the appearance of ferromagnetism in $\sigma_{xx}$-$x$ space but also in the magnetic field dependences of MCD, magnetization, and AHE. These results represent that the ferromagnetism is originated from the single source: most probably the charge carriers induce the ferromagnetism.

The authors gratefully acknowledge K. Ando for the access to MCD measurement system, K. Nakajima and T. Chikyow for TEM, T. Hasegawa, H. Koinuma, H. Ohno and F. Matsukura for discussions. This work was supported by the Japanese Ministry of Education, Culture, Sports, Science and Technology in Japan, Grant-in-Aid for Creative Science Research (14GS0204, 13NP0201) and for Young Scientists (A16686019), NEDO International Joint Research program (02BR3), and the Sumitomo Foundation.

H. T. is supported by Research Fellowship of the Japan Society for the Promotion of Science for Young Scientists.

**Figure captions**

FIG. 1. (Color online) The optical absorption (a) and MCD (b) spectra at room temperature for $Ti_{0.97}Co_{0.03}O_{2-\delta}$ films with various carrier concentrations ($n$). The magnetic field of 1 T is perpendicular to the film plane for (b).The inset shows the MCD spectra in various magnetic fields at room temperature for $n = 2 \times 10^{20}$ $cm^{-3}$ sample.

FIG. 2. (Color online) The MCD spectra in 1 T at room temperature for $Ti_{1-x}Co_xO_{2-\delta}$ with different $x$ grown in (a) $P_{O2} = 10^{-6}$ Torr and (b) $P_{O2} = 10^{-7}$ Torr. The magnetic field dependence of MCD at photon energies of (c) 2.50 eV and (d) 3.05 eV.

FIG. 3. Mapping in $\sigma_{xx}$-$x$ space for ferromagnetic (solid symbol) and paramagnetic (open symbol) responses determined from MCD experiments. Circle, square, triangle, and diamond symbols represent the samples grown in $P_{O2}$ of $10^{-7}$ Torr, $10^{-6}$ Torr, $10^{-5}$ Torr and $10^{-4}$ Torr, respectively. The appearance of AHE exactly coincides with this mapping.

FIG. 4. (Color online) Magnetic field dependence of MCD (red curve), Hall resistivity (blue curve), and magnetization (green symbol) at room temperature for $Ti_{0.90}Co_{0.10}O_{2-\delta}$ samples with (a) $n = 2 \times 10^{20}$ $cm^{-3}$ and (b) $4 \times 10^{21}$ $cm^{-3}$. The magnetic field is perpendicular to the film plane for all the measurements. The sign of MCD is inverted from that in Figs. 2 (c) and (d) for comparison.

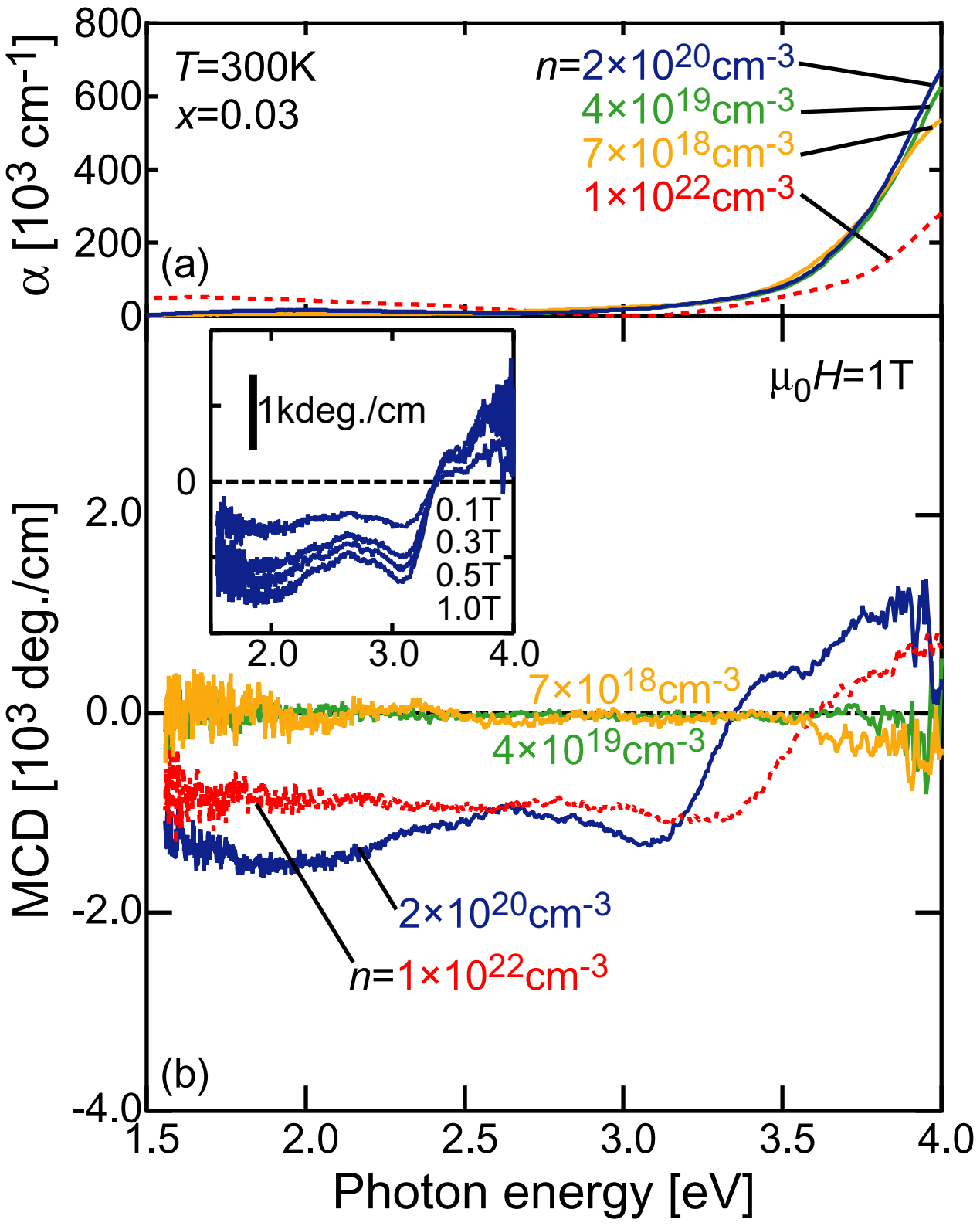

Fig.1. H. Toyosaki et al.

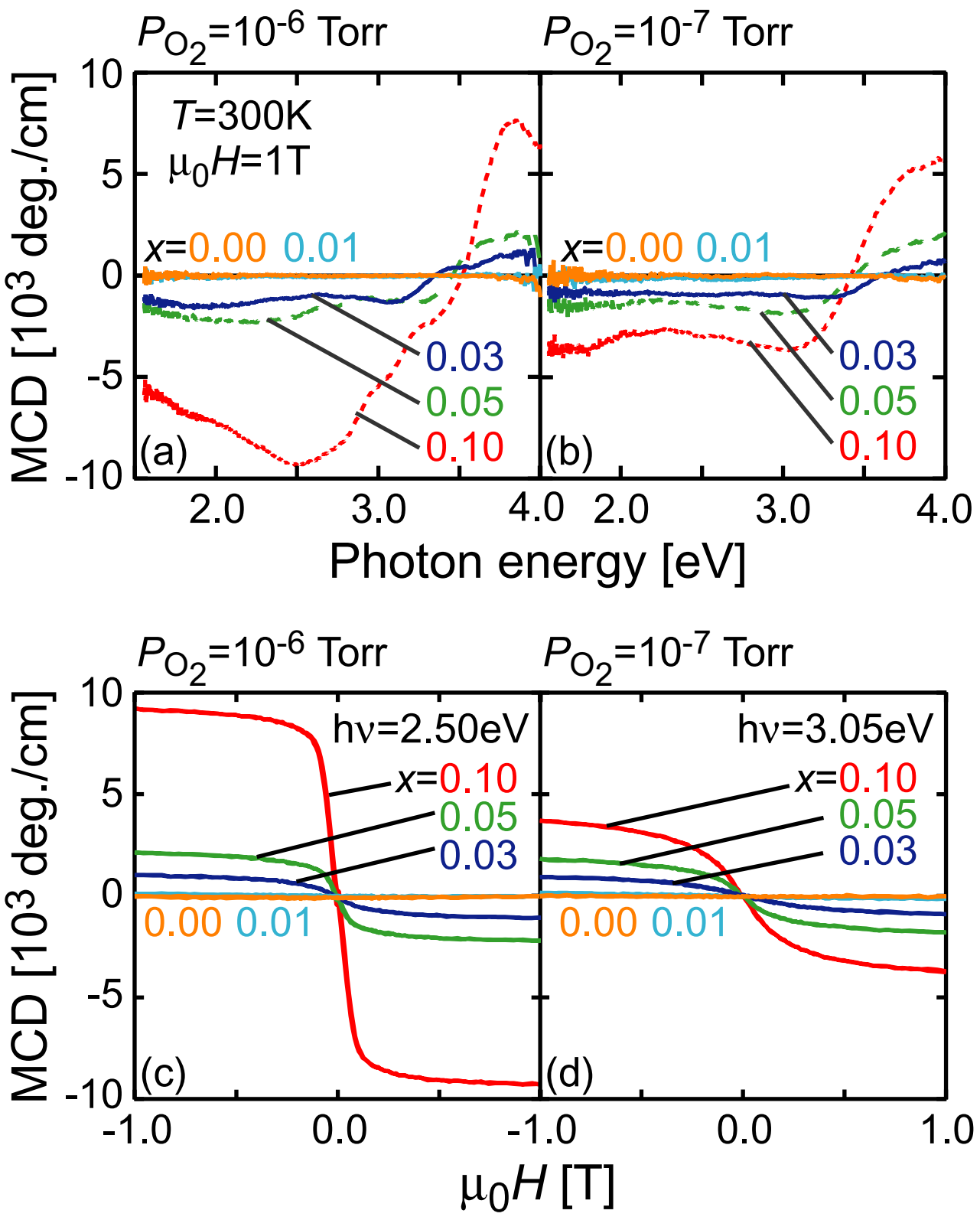

Fig.2 H. Toyosaki et al.

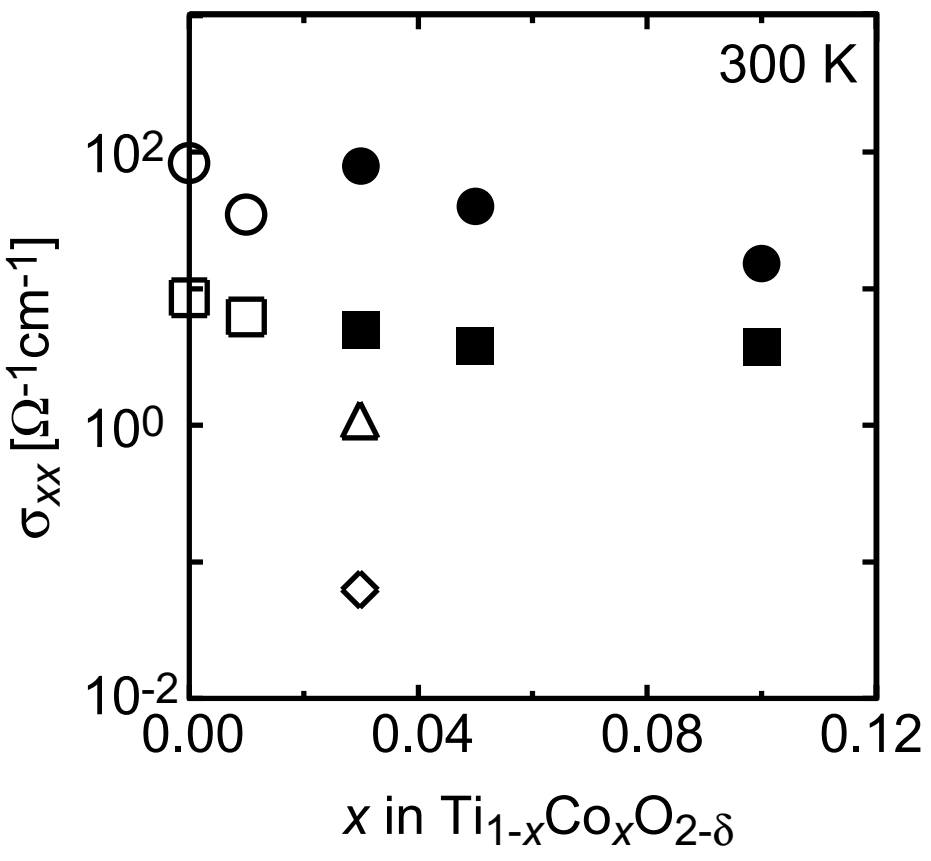

Fig.3. H. Toyosaki et al.

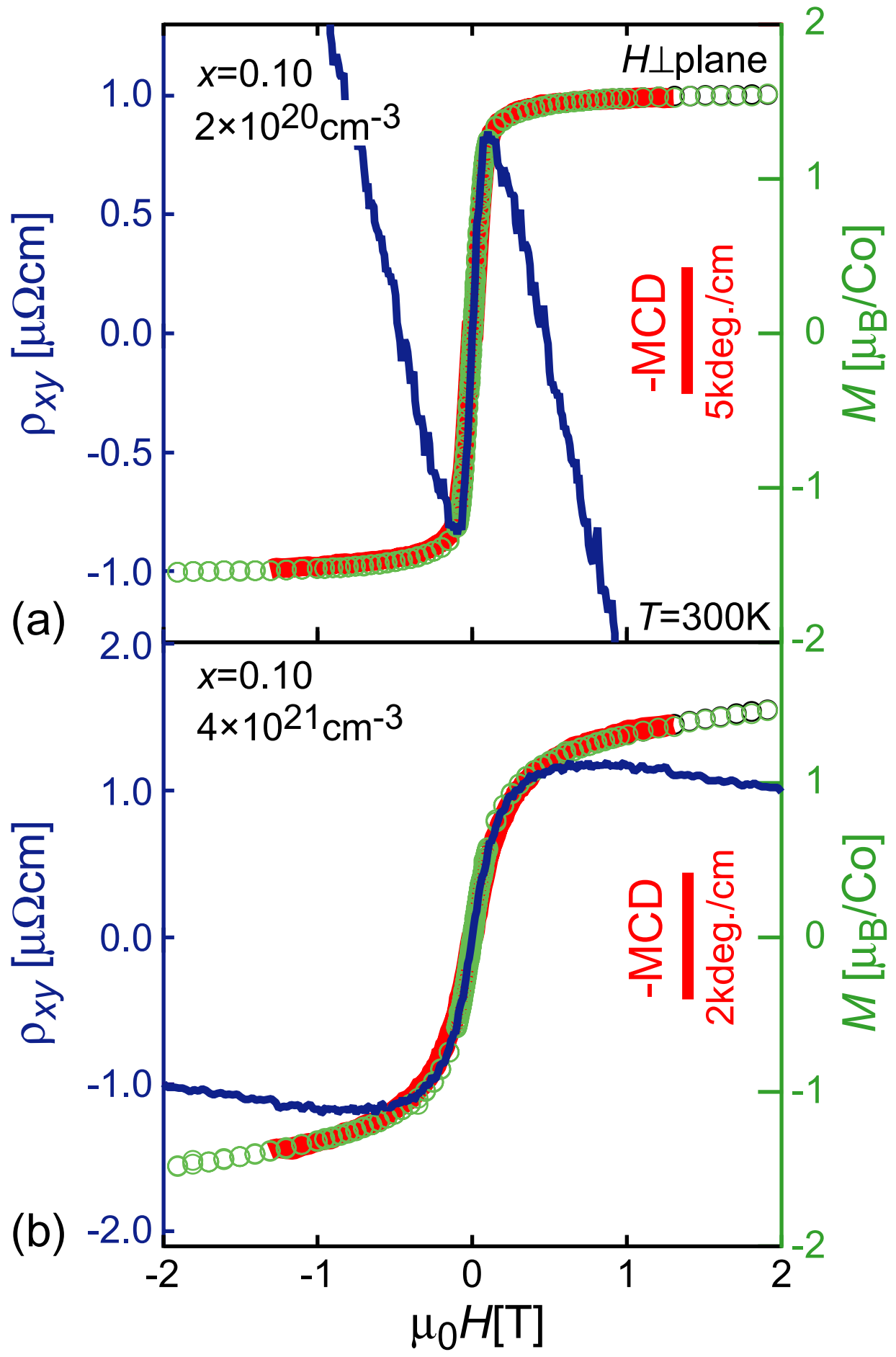

Fig.4 H.Toyosaki et al.